\documentclass[useAMS,usenatbib]{mn2e}
\usepackage{graphicx}
\usepackage{rotating}
\usepackage{lscape}

\title[Theoretical fit of Cepheids in NGC 1866]{Theoretical fit of
  Cepheid light an radial velocity curves in the Large Magellanic
  Cloud cluster NGC 1866.} 
\author[M. Marconi, R. Molinaro, V. Ripepi, I. Musella and
  E. Brocato]{M. Marconi$^{1}$\thanks{E-mail: marconi@na.astro.it},
  R. Molinaro$^{1}$, V. Ripepi$^{1}$, I. Musella$^{1}$ and E. Brocato$^{2}$\\ 
$^{1}$INAF-Osservatorio Astronomico di Capodimonte, Salita
  Moiariello, 16, 80131, Napoli, ITALY\\
$^{2}$INAF-Osservatorio Astronomico di Roma, Via Frascati, 33, 00040 Monte Porzio Catone, ITALY\\}
\begin{document}

\date{Accepted ..... Received .... ;}

\pagerange{\pageref{firstpage}--\pageref{lastpage}} \pubyear{2002}

\maketitle

\label{firstpage}

\begin{abstract}
We present a theoretical investigation of multifilter (U,B,V, I and K)
light and radial 
velocity curves of five Classical
Cepheids in NGC 1866, a young massive cluster of the Large Magellanic
Cloud.  The best fit models accounting for the luminosity and radial
velocity variations of the five selected variables, four pulsating in
the fundamental mode and one in the first overtone, provide direct
estimates of their intrinsic stellar parameters and individual
distances.
The resulting stellar properties indicate a slightly brighter
Mass--Luminosity relation than the canonical one, possibly due to mild
overshooting and/or mass loss. As for the inferred distances, the
individual values are consistent within the
uncertainties. Moreover, their weighted mean value corresponds to a distance
modulus of 18.56$\pm$0.03 (stat) $\pm$0.1 (syst) mag, in agreement with several independent
results in the literature.
\end{abstract}

\begin{keywords}
star clusters -- NGC 1866 -- Cepheids -- Variable stars .
\end{keywords}

\section{Introduction}
Classical Cepheids are considered the most important primary distance indicators
within the Local Group. Their
Period--Luminosity relation, discovered 
by Miss Leavitt in 1912 \citep[][]{l1912} for Cepheids in the
Small Magellanic Cloud and usually calibrated in the Large Magellanic
Cloud (LMC) \citep[see e.g.][]{mf91,u99}, is now at the basis of an
extragalactic distance scale \citep[see e.g.][and references
therein]{f01,s01}. Indeed, with the capabilities of the Hubble Space
Telescope, Cepheids have been observed at distances (up to $\sim$ 30
Mpc), enabling the calibration of several secondary distance
indicators capable to reach cosmological distances and to provide an
estimate of the 
Hubble constant $H_0$ \citep[see e. g.][for  detailed discussion]{f01}.
In spite of the most recent relevant efforts in the direction of reducing
the uncertainty on the Cepheid based extragalactic distance scale
\citep[see e.g.][and references therein]{rie11,rie12}, some systematic
effects, including the 
effect of the host galaxy metal content, remain unsolved,
and different authors keep to provide significantly
different   
estimates of the Hubble constant \citep[see e.g.][and references
  therein]{tr12, rie12}. 
The first crucial step for the calibration of the
extragalactic distance scale is the distance to the LMC. Several
methods have been adopted in the literature 
\citep[see e.g.][and references therein]{w11,mol12}
providing values that range from $\sim$ 18.1 to $\sim$ 18.9
mag. Systematic effects such as a non negligible metallicity dispersion,
 differential reddening and a significant depth of the Cloud,  are
 known to be at work.  

In this context, classical Cepheids belonging to young stellar
clusters in the LMC  play an important role, being at the same
distance and sharing the same age and chemical composition. Thanks to
these advantages they offer a unique opportunity to investigate the
uncertainties affecting both empirical approaches and theoretical
scenarios.  
NGC 1866 is one of the most massive young clusters in the age range 100-200
Myr, and it has been the subject of a very long list of papers,
starting with the pioneering ones by \citet{at67} and
\citet{r74}. Subsequent authors focused on studying the cluster either
as a testbed of stellar evolution theory
\citep[e.g.][]{br89,br94,br03,br04,ba02,w01,c89,t99},
as a Cepheid host \citep{wel91,g94,ws93,w95,gie00,sto05,tes07,mol12}, or
as a dynamical laboratory \citep[][]{fi92}. It has also been the
subject of a strong debate over the presence of convective
overshooting in intermediate-age stellar models, and on the fraction
of binaries in the main sequences. Moreover, this cluster lies in the
outskirts of the LMC, so that field contamination is not severe. 
The investigation of Cepheid properties in this cluster can provide
crucial information for our understanding of the physics and the
evolution of intermediate mass stars. In particular the comparison with
the predictions of pulsation properties based on hydrodynamical models
is an important tool to constrain the individual distances and the
intrinsic stellar parameters, without relying on stellar evolution
models, and in turn independently constraining their physical and
numerical assumptions. In particular, the possibility to obtain this information from the direct comparison between modeled and observed light curves has been first claimed  by
\citet{was97} for Classical Cepheids and by \citet{bcm00} for RR Lyrae.
Subsequent applications both to field and cluster pulsating stars, sometimes including the additional match of radial velocity or radius curves, have
provided self-consistent results, also in agreement with independent
estimates in the literature \citep[see][]{dif02,bcm02,mc05,md07,nat08,m10}.
 
In this paper we present an accurate comparison between observed
multifilter light and radial velocity curves for a sample of Cepheids
in NGC1866 and the theoretical counterparts based on nonlinear
convective models. The photometric and radial velocity data are
introduced in Sec.\ref{sec-data}, while the fitting procedure is
described in Sec.\ref{sec-fit}. The results of our analysis are
contained in Sec.\ref{sec-results} and include the best fit structural
parameters, the comparison with spectroscopic data, the implications for
the distance, including a critical discussion of the associated uncertainty, and the Mass--Luminosity relations. Finally,
Sec.\ref{sec-conclusions} contains the conclusions of the paper.

\section{The data}\label{sec-data}
The adopted photometric data include observations in
the U, B, V, I bands \citep[][Musella et al. in preparation]{mus06}
with the addition of the near 
infrared K band \citep{sto05, tes07}. To properly sample the region
near the maximum of light, we have integrated the B, V, I photometry of the
Cepheids HV 12198 and HV 12199 using the observations from
\citet{wel91}, MACHO\footnote{http://macho.anu.edu.au/} and
\citet{gie00} . Moreover, in some cases we excluded one photometric
band because the light curve was poorly sampled and/or lacking the
maximum and/or the minimum phases.  

To compare the Near Infrared data with models we have transformed the K band
measurements from CIT and LCO into the Johnson photometric system,
using the
relations by \citet{bes88}: 
\begin{equation}
K_J = K_{CIT/LCO}+0.027-0.007(V-K_{CIT/LCO}).
\end{equation}

As for the radial velocities, we used data from \citet{sto05},
\citet{sto04}, \citet{wel91} and \citet{mol12}.   

Table \ref{tab-data} summarizes the adopted number of photometric and radial
velocity measurements for the selected Cepheids.   

We phased the light curves by requiring that the B band
maximum of light occurred at phase zero,  while the maximum in the
other bands were shifted in phase as expected \citep[see
  e.g.][]{lab97,fre88}. Similarly, the radial velocity curves were
  phased by requiring that their minimum occurred at phase zero.

\begin{table}
\begin{center}
\caption{The period P, the apparent V magnitude and the number of
  measures for the U, B, V, I, K light curves and radial velocity
  curves of all selected Cepheids.}  
\begin{tabular}{c @{} c @{} c @{} c @{} c @{} c @{} c @{} c @{} c}
\hline
\hline
Name &\hspace{0.1 cm}Period &\hspace{0.1cm} V(mag) &\hspace{0.1 cm} U
&\hspace{0.1cm} 
B &\hspace{0.1cm} V &\hspace{0.1cm} I &\hspace{0.1cm} K
&\hspace{0.1cm} Rad. Vel.\\
\hline
HV 12197 &\hspace{0.1 cm} 3.143742 &\hspace{0.1cm} 15.91 &\hspace{0.1
  cm} 3 &\hspace{0.1 cm} 
69 &\hspace{0.1 cm} 87 &\hspace{0.1 cm} 38 &\hspace{0.1 cm} 35
&\hspace{0.1 cm} 38\\ 
HV 12198 &\hspace{0.1 cm} 3.522805 &\hspace{0.1cm} 15.77 &\hspace{0.1
  cm} 30 &\hspace{0.1 
  cm}  69 &\hspace{0.1 cm} 90 &\hspace{0.1 cm} 62 &\hspace{0.1 cm} 77
&\hspace{0.1 cm} 38\\ 
HV 12199 &\hspace{0.1 cm} 2.639181 &\hspace{0.1cm} 16.09 &\hspace{0.1 cm} 34 &\hspace{0.1
  cm}  157 &\hspace{0.1 cm} 199 &\hspace{0.1 cm} 172 &\hspace{0.1 cm}
54 &\hspace{0.1 cm} 39\\ 
We 2 &\hspace{0.1 cm} 3.054847 &\hspace{0.1cm} 15.86 &\hspace{0.1 cm} 15 &\hspace{0.1 cm} 69
&\hspace{0.1 cm} 90 &\hspace{0.1 cm} 62 &\hspace{0.1 cm} 5
&\hspace{0.1 cm} 12\\
V 6 &\hspace{0.1 cm} 1.944252 &\hspace{0.1cm} 15.97 &\hspace{0.1 cm} 21 &\hspace{0.1 cm} 69
&\hspace{0.1 cm} 90 &\hspace{0.1 cm} 62 &\hspace{0.1 cm} 10
&\hspace{0.1 cm} 10\\
\hline
\hline
\end{tabular}
\label{tab-data}
\end{center}
\end{table}

\section{Model fitting}\label{sec-fit}
New nonlinear convective pulsation models have been computed to
reproduce the observed luminosity and radial velocity variations. 
The adopted theoretical framework is based on a nonlinear radial
pulsation code, including the non-local and time-dependent treatment of
turbulent convection \citep{s82,bs94,bms99}. The system of nonlinear
equations is closed using a free parameter, $\alpha_{ml}$, that is
proportional to the mixing-length parameter. Changes in the mixing
length parameter affect, as expected, both the limit cycle stability
(pulsation amplitudes) and the topology of the instability strip
\citep{f07,dic04}. Similar approaches for the treatment of convective
transport have been developed by \citet{feu99,busz07,ow05}. 
We remind here that the nonlinearity and the inclusion of a non
local, time--dependent treatment of convection and of its coupling with
pulsation allows us to accurately predict all the relevant observables
of stellar pulsation, namely the complete topology of the instability
strip for each selected pulsation mode, the accurate morphology of
light, radial velocity and radius variations and the associated
pulsation amplitudes. 

The modeling of the observed light and radial velocity curves has been
organized in three main steps:\begin{itemize}

\item First, we constructed a set of models with fixed chemical
composition (Z=0.008, Y=0.25, consistent with the abundances measured
for Cepheids in NGC 1866 \citep{muc11}),  mass and period (equal to
the observed one), varying the effective temperature and luminosity
in order to reproduce the observed variations.  

\item Once identified the best effective temperature from the previous
 step, we built a sequence of models at fixed chemical composition,
 period and effective temperature, by varying the mass and the
 luminosity in order to obtain the best fit model, reproducing
 simultaneously the multifilter light curves and the radial velocity one.

\item In some cases we also changed slightly the metal and helium
  abundance within current uncertainties on the LMC chemical composition.

\end{itemize}

Each model curve was phased in order to find the
maximum of light in the B band  at phase zero. 
Afterward, for each model we calculated the shifts in magnitude and
phase, which gave the best match between the theoretical light curves
and the observational data. 
Specifically, for each model, we  calculated the phase shift,
$\delta \phi$, and the magnitude shift, $\delta M$, which minimized the
following $\chi^2$ function: 
\begin{equation}\label{eq-chi2-phot}   
\chi^2=\sum_{i=0}^{N_{band}}\sum_{j=1}^{N_{points}}\left [m^i_j -\left 
(M^i_{mod}\left (\phi^i_j+\delta \phi^i\right )+\delta M^i\right )\right ]^2
\end{equation}
where the index i runs over the number of photometric bands,
$N_{band}$, and j over the number of measurements, $N_{points}$. In the
equation above we used a spline interpolation to evaluate the
theoretical absolute magnitude $M_{mod}$ at the phase $\phi_j$, of the
jth photometric measurement $m_j$, plus the shift $\delta \phi$.
We note that the parameter $\delta M^i$ gives the distance modulus of
the analyzed Cepheid in the i$_{th}$ photometric band. 

As for  the radial velocity curves, we transformed the model
pulsational velocity, $v^{puls}$, into radial velocity, 
$v^{rad}$, by using the projection factor $p$ in the relation
$v^{rad}=-\frac{1}{p} v^{puls}$ and phased it in order to find  its minimum
value at phase zero. In the previous formula we fixed the
projection factor to p=1.27, obtained by using the mean period for the
selected Cepheids  \citep[first-overtone fundamentalized
  according to][]{fea97} in the equation $p=1.31-0.08\log P$, by
\citet{nar09}.

Then, to match the model radial velocity variations with the
data, we minimized the $\chi^2$ function of an equation similar to eq.(\ref{eq-chi2-phot}), where
the magnitude shift is replaced with the $\gamma$ barycentric
velocity, namely: 
\begin{equation}\label{eq-chi2-vr}   
\chi^2=\sum_{j=1}^{N_{points}}\left [v^{rad}_j -
\left (-\frac{1}{p}v^{puls}\left (\phi_j+\delta \phi\right
)+\gamma \right )\right ]^2 
\end{equation}

\section{Results}\label{sec-results}
In the present section we discuss the results obtained from the
fitting procedure. The match between the best fit models and the data
is described for all the selected Cepheids. The derived structural
parameters for all Cepheids are given and for some of them we
performed a comparison with the results obtained from the
spectroscopy. Finally, we will give the best fit distance moduli and 
reddening values and discuss the implications of the derived
Mass--Luminosity relation. 

\subsection{Best fit models}
For each Cepheid we identified a best fit model and other four models
(hereafter ``secondary'' models) which are the closest (in the sense
of the value of the $\chi^2$ function) to the best fit one, according
to a reasonable\footnote{Smaller steps do not produce significant
  differences in the model properties and/or are within the numerical
  precision of the hydrodynamical code.}
  selection of the model grid steps (typically 25 K in
effective temperature and 0.02 dex in $\log{L/L_{\odot}}$). Two of
these ``secondary'' models have the same temperature of the best fit
one and a varied value of the luminosity (and the mass), and, vice
versa, the other two models have a fixed value of the luminosity and a
varied value of  the temperature. These secondary models are used to
define the errors on the parameters. In particular, we defined the
$\pm 1 \sigma$ uncertainty interval as half of the difference between
the parameters of the quoted ``secondary'' models and the ones of the
best fit model. Typically this corresponds to errors of $\sim \pm 12
K$, $\sim \pm 0.1 M_{\odot}$ and $\sim \pm 0.01$ dex in effective
temperature, mass and luminosity respectively. 

To analyze the effect of the projection factor $p$ on the fit of
the radial velocity data, we also tried to vary it in the $\chi^2$
function given by Eq.\ref{eq-chi2-vr}. The resulting best fit values
of the p factor are listed in Tab.\ref{tab-params} together with all
the other structural parameters derived from our analysis.

We also computed additional models varying the chemical composition
but these tests did not improve the accuracy of the model fitting for
none of the investigated Cepheids, suggesting the usually assumed  
Z=0.008 for the metallicity of these stars.
The results in Tab.\ref{tab-params} also show that the value of the
assumed mixing length parameter  (to close the nonlinear system of
dynamical and convective equations) that provides the best match is 
$\alpha_{ml}=1.9-2.0$ for fundamental variables and a slightly smaller value
($\alpha_{ml}=1.8$) for the first overtone one, in agreement with
previous theoretical results based on the analysis of both Cepheid and
RR Lyrae properties \citep[see e.g.][and references
  therein]{bcm02,dic04,mc05,f07,nat08}.  

Below, we describe the result of the fit of light and radial
velocity curves for all the selected Cepheids.

\subsubsection{We 2}
\begin{figure*}
\begin{center}
\includegraphics[width=180mm]{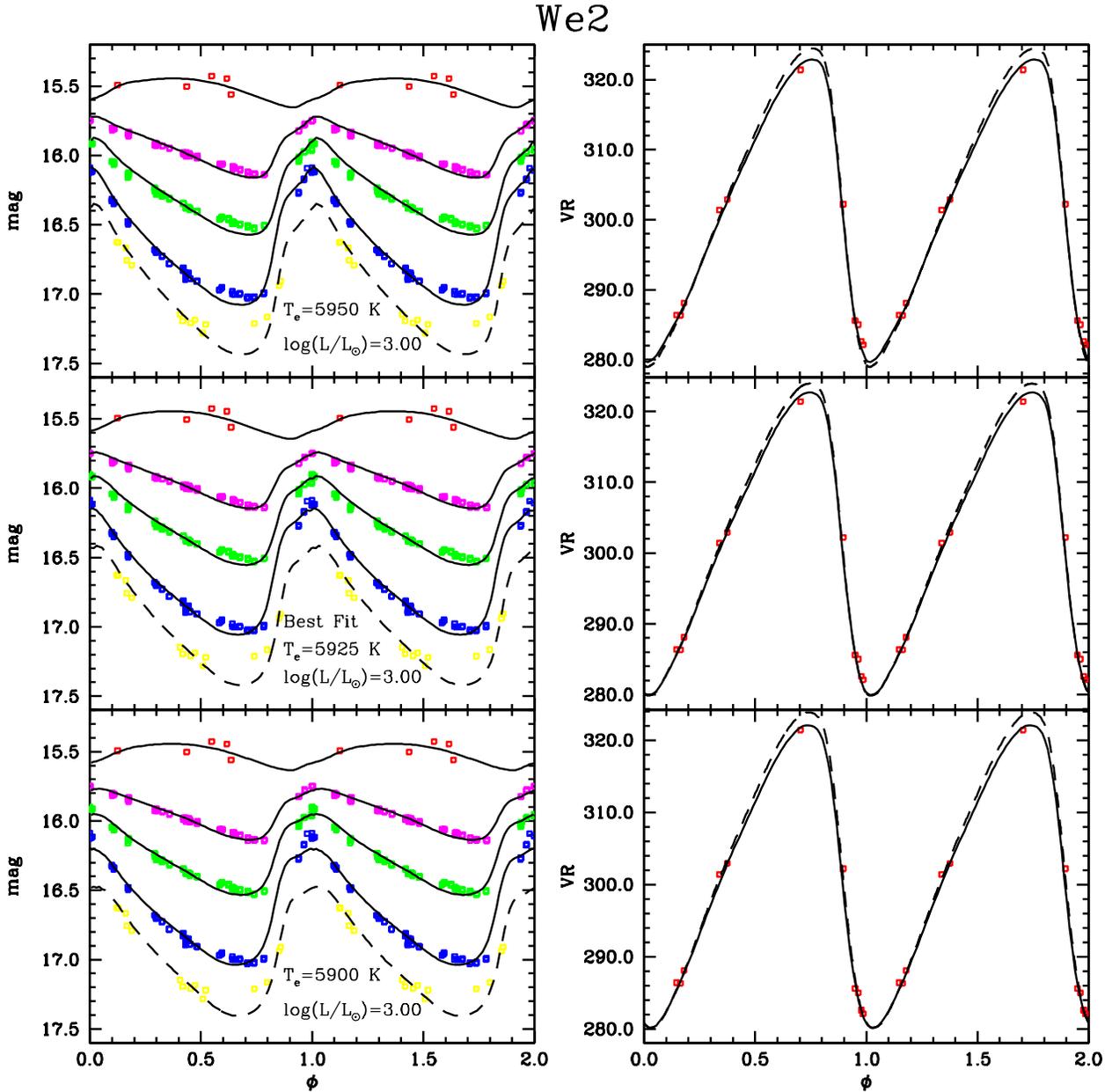}
\caption{In each one of the three panels on the left, the U, B, V, I
  and K band observed light curves of the Cepheid We 2 are shown from
  bottom to top (empty squares). The data have been systematically
  shifted to be shown in the same plot. The radial velocity data of
  the same star is shown in the three panels on the right (empty
  squares). The solid lines represent the model  matched to the
  data. The U band model is plotted with dashed line because it will
  be excluded in our further analysis (see Sec.\ref{sec-dist}). The best
  fit (BF) is plotted in the central panels and the models 
  corresponding to $T_e^{BF}-2\delta T$ and $T_e^{BF}+2\delta T$, are
  shown in the bottom and top panels 
  respectively. In the panels showing the radial velocity curves, the
  models with p-factor free to vary are also plotted (long dashed
  lines).}\label{fig-bf-temp-we2}                
\end{center}
\end{figure*}

\begin{figure*}
\begin{center}
\includegraphics[width=180mm]{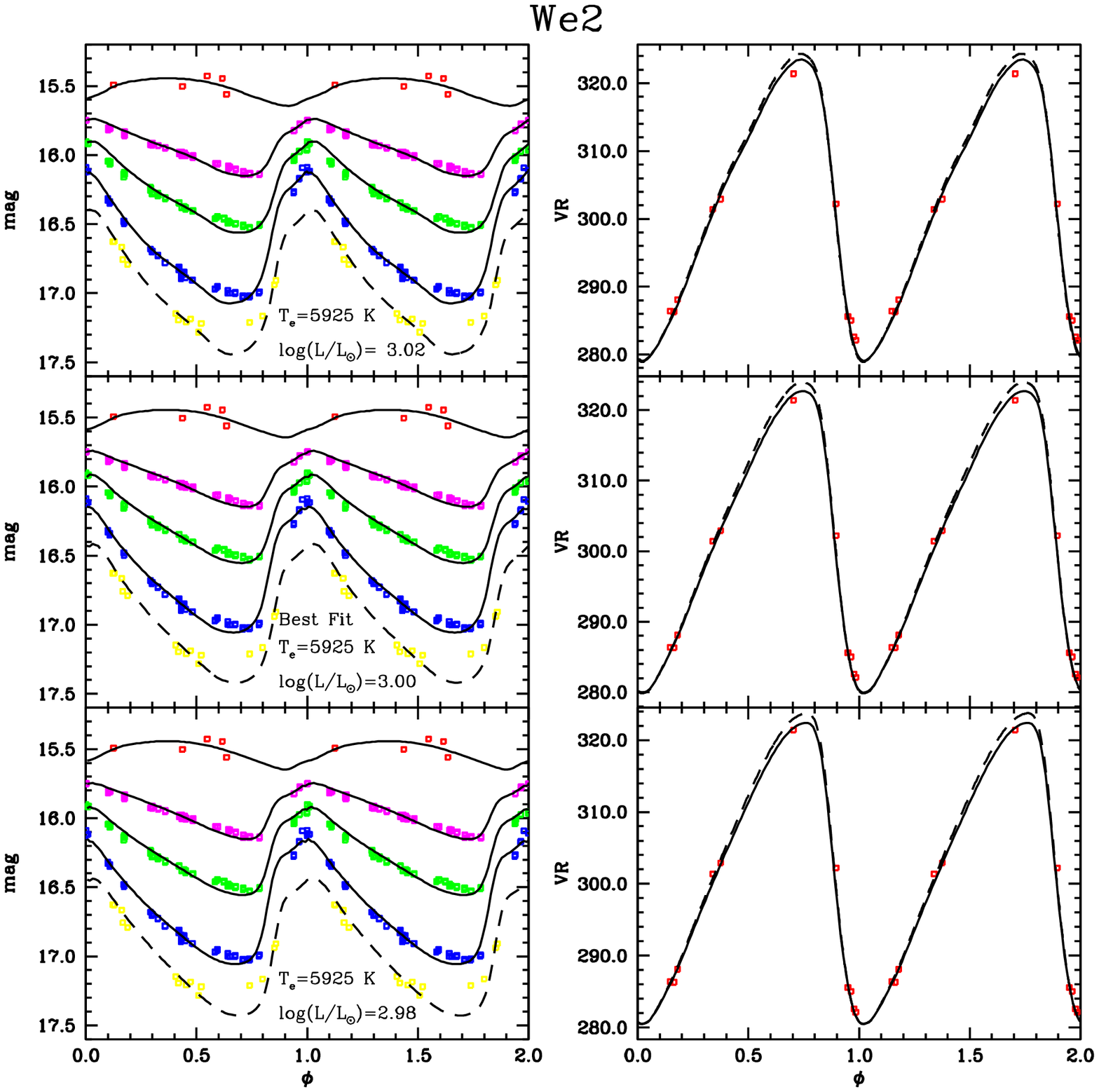}
\caption{The same as Fig.~\ref{fig-bf-temp-we2} but with models
  corresponding to $\log(L/L_\odot)^{BF}-2\delta (\log L/L_\odot)$ and $\log
  (L/L_\odot)+2\delta \log(L/L_\odot)$, for
  We 2. } \label{fig-bf-llsun-we2}
\end{center}
\end{figure*}

The best fit model selection illustrated above is shown for the fundamental pulsator
We 2 in Figs.\ref{fig-bf-temp-we2}-\ref{fig-bf-llsun-we2}. The best
fit model (central panels in the two figures) has a characteristic effective
temperature of 5925 K and a luminosity  $\log(L/L_\odot)=3.00$ dex. As evident
in the top and bottom panels, a variation of 25 K in the temperature
(at fixed luminosity) or 0.02 dex in the luminosity (at fixed
temperature) worsen the match with the data. As mentioned above,
we  define the parameter uncertainty intervals as the half of the
quoted variations. We also note that the chosen value of the projection
factor p=1.27 provides an excellent match of the model
radial velocity with the plotted data, even if the best analytic match is
obtained for p=1.23. 

\subsubsection{V6}
\begin{figure*}
\begin{center}
\includegraphics[width=160mm]{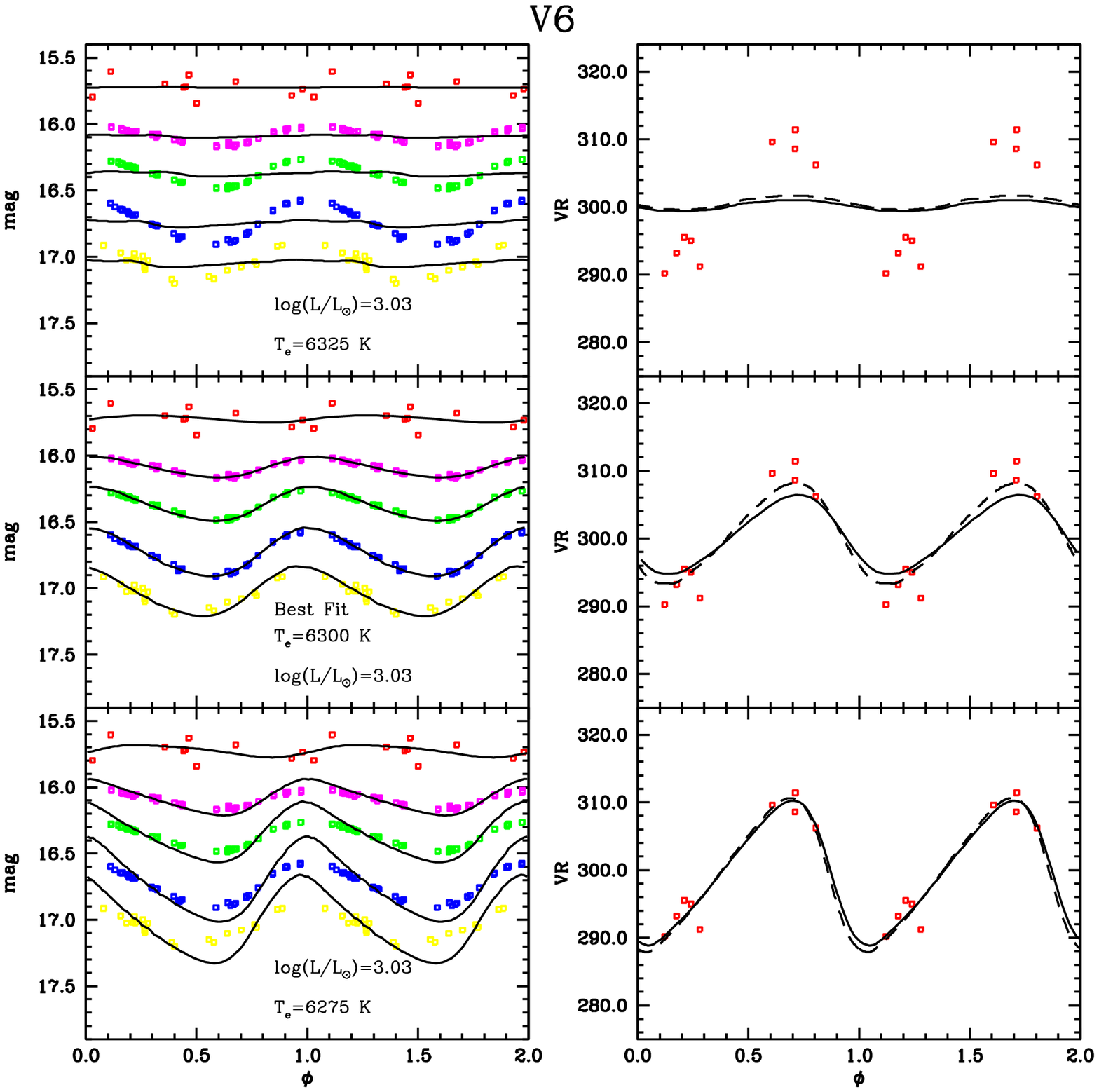}
\caption{The same as Fig.~\ref{fig-bf-temp-we2} for
  V6.}\label{fig-bf-temp-v6}      
\end{center}
\end{figure*}

\begin{figure*}
\begin{center}
\includegraphics[width=160mm]{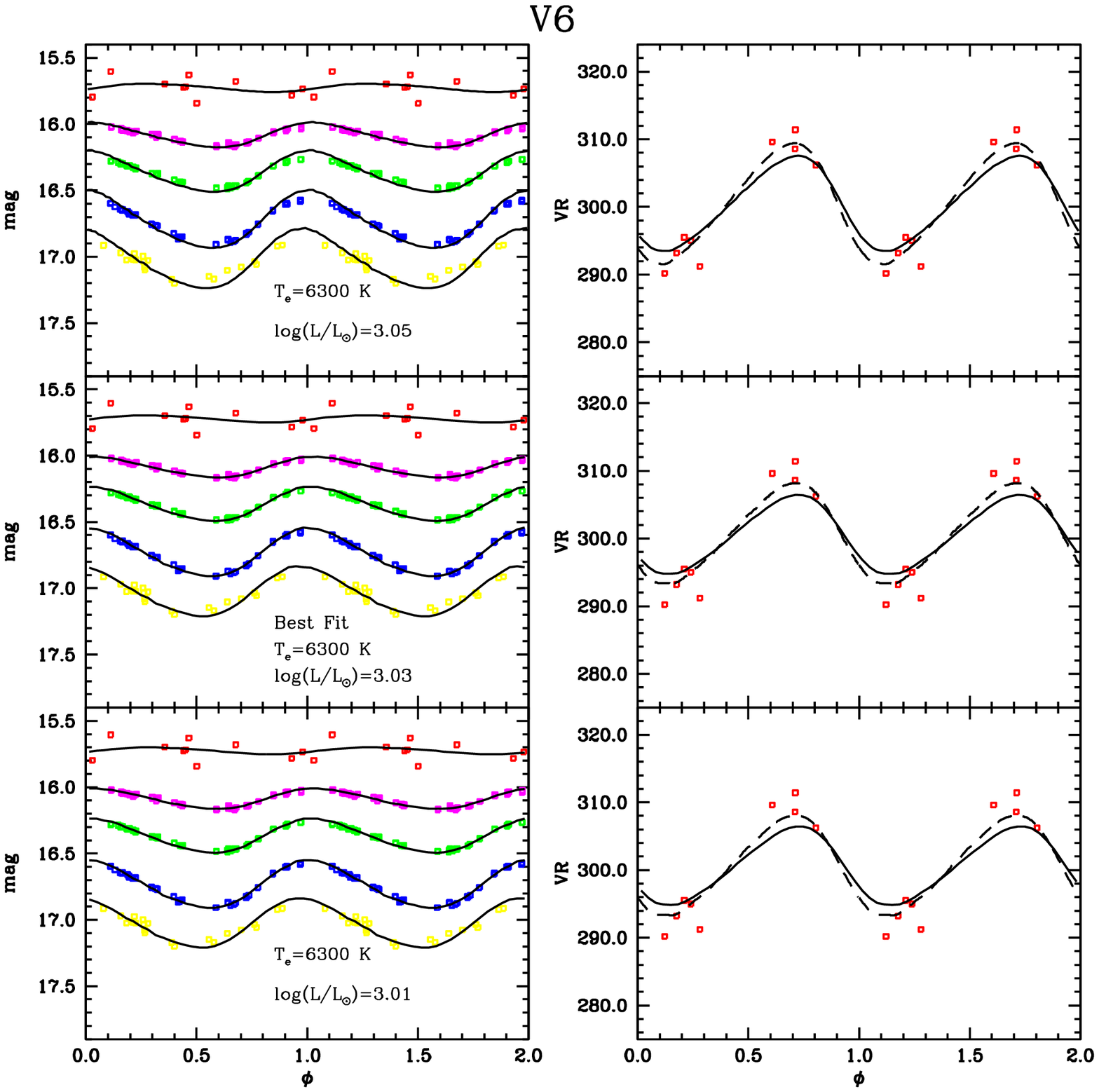}
\caption{The same as Fig.~\ref{fig-bf-llsun-we2} for
  V6.}\label{fig-bf-llsun-v6}      
\end{center}
\end{figure*}

It is instructive to show the same plots than We 2 for the first
overtone pulsator V6 (Figs.\ref{fig-bf-temp-v6}-\ref{fig-bf-llsun-v6}). 
In this case we note that the best fit model is located very close to
the first overtone blue edge, so that a shift of only $25 K$ towards
higher effective temperatures (see the top panel of
Fig.\ref{fig-bf-temp-v6}) almost quenches the pulsation.
On the other hand when decreasing the effective temperature by $25 K$
(bottom panel of the same figure), the predicted amplitudes increase
significantly beyond the observed ones. If we consider the effect of a
variation in the stellar mass we note that the corresponding variation
in the pulsation amplitudes is much smaller, even if non negligible
when computing the  $\chi^2$ minimization.

The procedure of best fitting of the radial velocity curve, when
considering the projection factor as a free parameter, provides the
value p=1.00, which is significantly different from 1.27 and smaller
than other typical values adopted in the literature. A possible
explanation might be  a not sufficiently  good quality of the data,
but we note that similar results have been discussed by
\citet{nat08} for $\delta$ Cephei using a well sampled radial velocity
curve.   
 
\subsubsection{HV 12197}
\begin{figure*}
\begin{center}
\includegraphics[width=110mm]{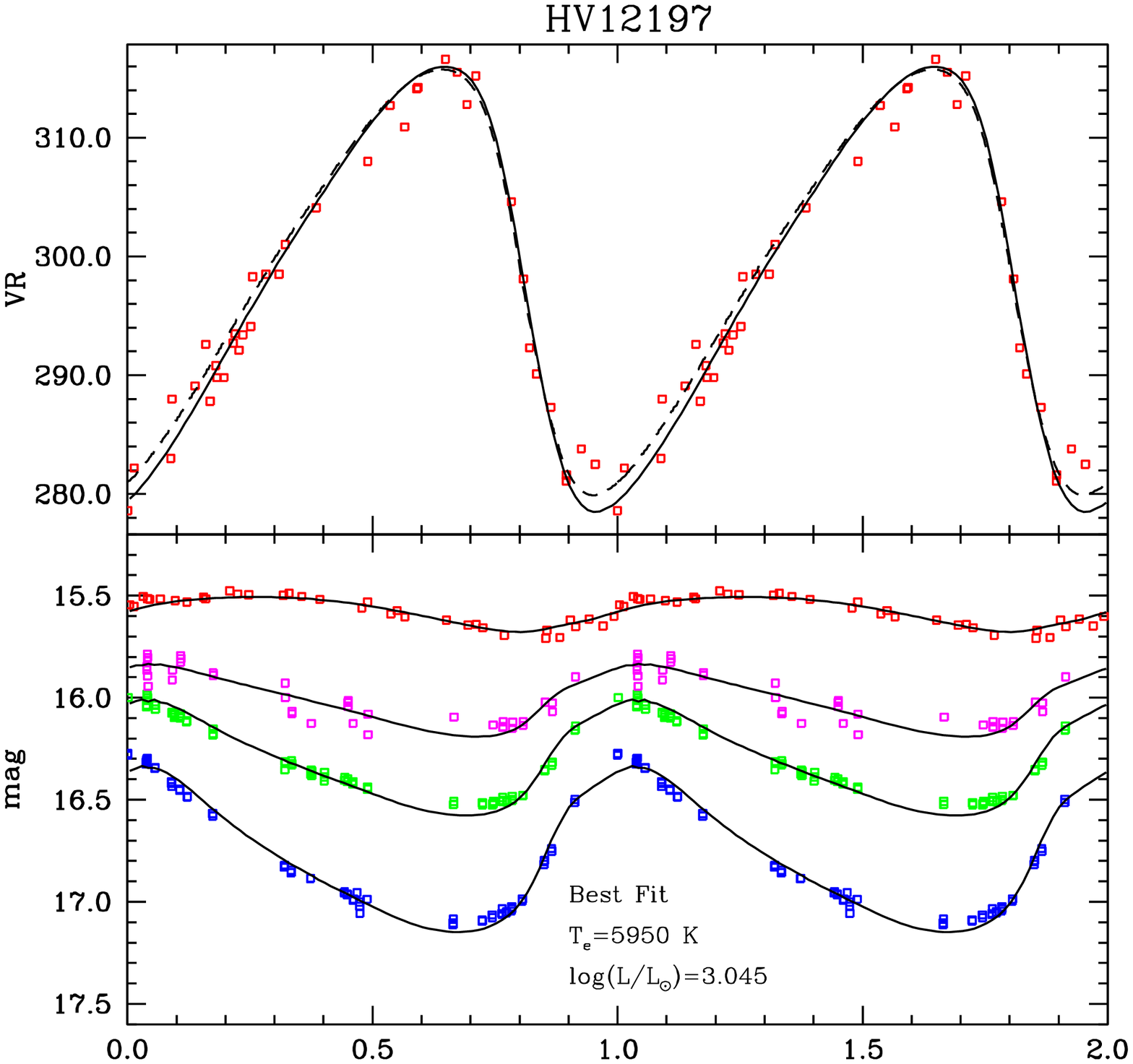}
\caption{The best fit model light curves and the model radial velocity
  curves (solid lines) for the Cepheid HV 12197 are matched with the
  data (empty squares), in the bottom and top panel respectively. The
  plotted light curves are in the B, V, I and K bands from bottom to
  top. The best fit radial velocity curve obtained by varying the
  projection factor is shown as dashed line.}\label{fig-bf-hv12197}        
\end{center}
\end{figure*}

The two panels of Fig.\ref{fig-bf-hv12197} show the best fitting result
for Cepheid HV 12197. The U band is excluded from our analysis due to
the poor light curve. According to the best fit model this star is
the most luminous of our sample ($\log(L/L_\odot)=3.045$ dex) and has an
effective temperature of 5950 K. As for the radial velocity curve, the chosen
value of the projection factor (1.27) provides a good match with the data and
it is not significantly different from the value p=1.33 obtained from
the $\chi^2$ minimization with variable p.

\subsubsection{HV 12199}
\begin{figure*}
\begin{center}
\includegraphics[width=180mm]{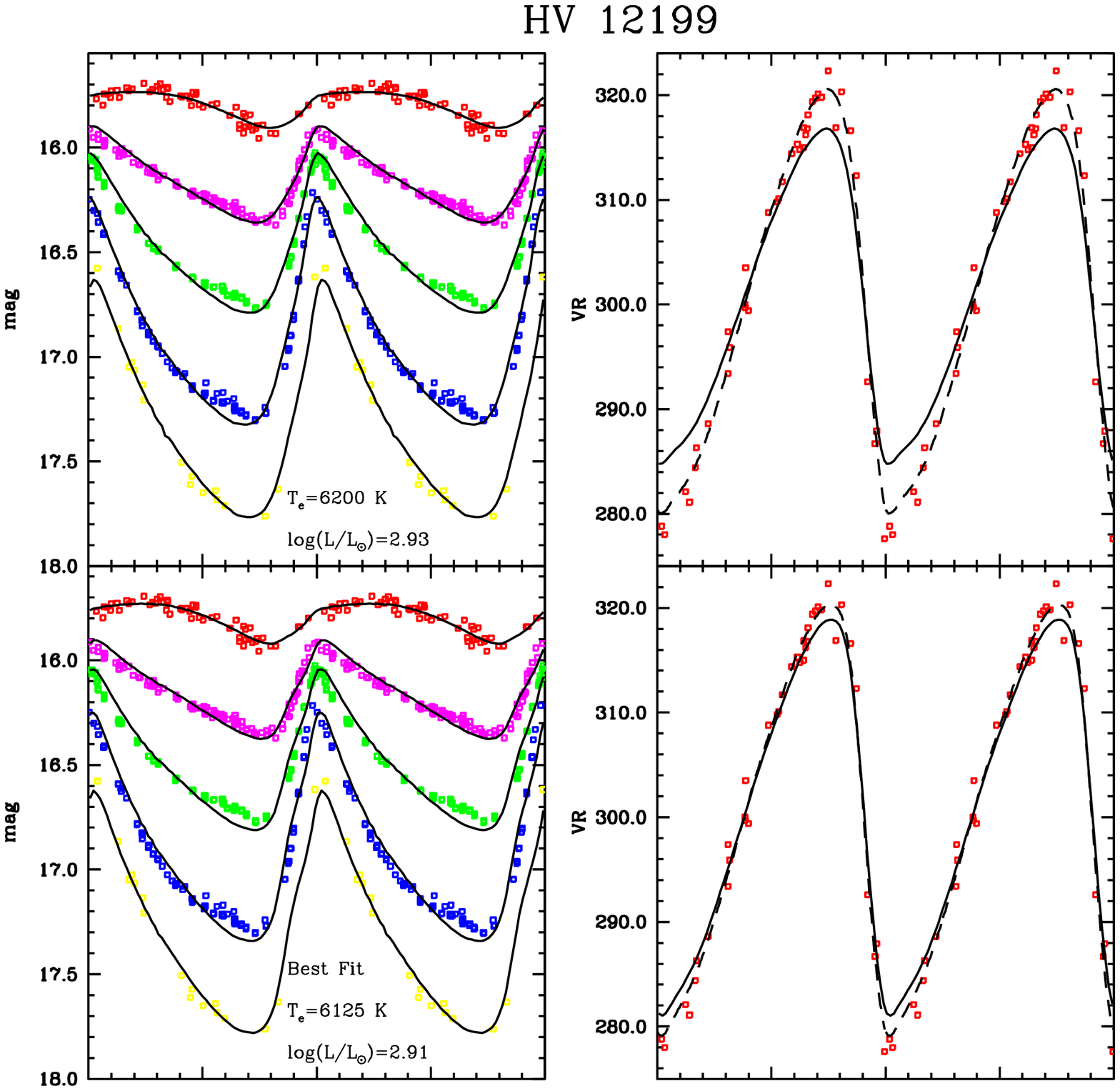}
\caption{In the bottom panels the match between the best fit model and
  data, for the Cepheid HV 12199, is plotted with the same
  meaning of the used symbols than in the
  Fig.\ref{fig-bf-hv12197}. The plots in the top panels refer to a
  model with T=6200 K, $\log(L/L_\odot)=2.93$ and $M/M_\odot=3.5$.
}\label{fig-bf-hv12199}        
\end{center}
\end{figure*}

The best fit model for the Cepheid HV 12199 is shown in the bottom
panels of Fig.\ref{fig-bf-hv12199}. The luminosity derived from the
fitting procedure ($\log(L/L_\odot)=2.91$ dex) results to be the smallest
in the selected sample and the effective temperature of this star is
6125 K. The model radial velocity curve, plotted in the right panel,
shows a small  discrepancy in its amplitude with the data if we fix the
value of the p factor to 1.27. It is necessary to decrease the
projection factor to 1.17 in order to achieve the best match. It is
interesting to note that, if we focus our attention only on the 
photometric data, the resulting best fit model for HV 12199 is given in the
top panels of Fig.\ref{fig-bf-hv12199}, with a predicted effective
temperature and $\log(L/L_\odot)$ of 6200 K and 2.93 dex,
respectively. However, we excluded this model because the radial
velocity curve, obtained from the chosen value of the projection
factor, has a too small amplitude to fit  the data, (see the solid line in the
Fig.\ref{fig-bf-hv12199}). To account for the amplitude of the radial
velocity data it would be necessary to decrease the projection factor
to the too small value p$\sim$1.0 (dashed line in the Figure).   
 This inconsistency that can be indeed due to a too high model
  effective temperature, led us to prefer the best model reported in 
Fig.\ref{fig-bf-hv12199}

\subsubsection{HV 12198}
\begin{figure*}
\begin{center}
\includegraphics[width=180mm]{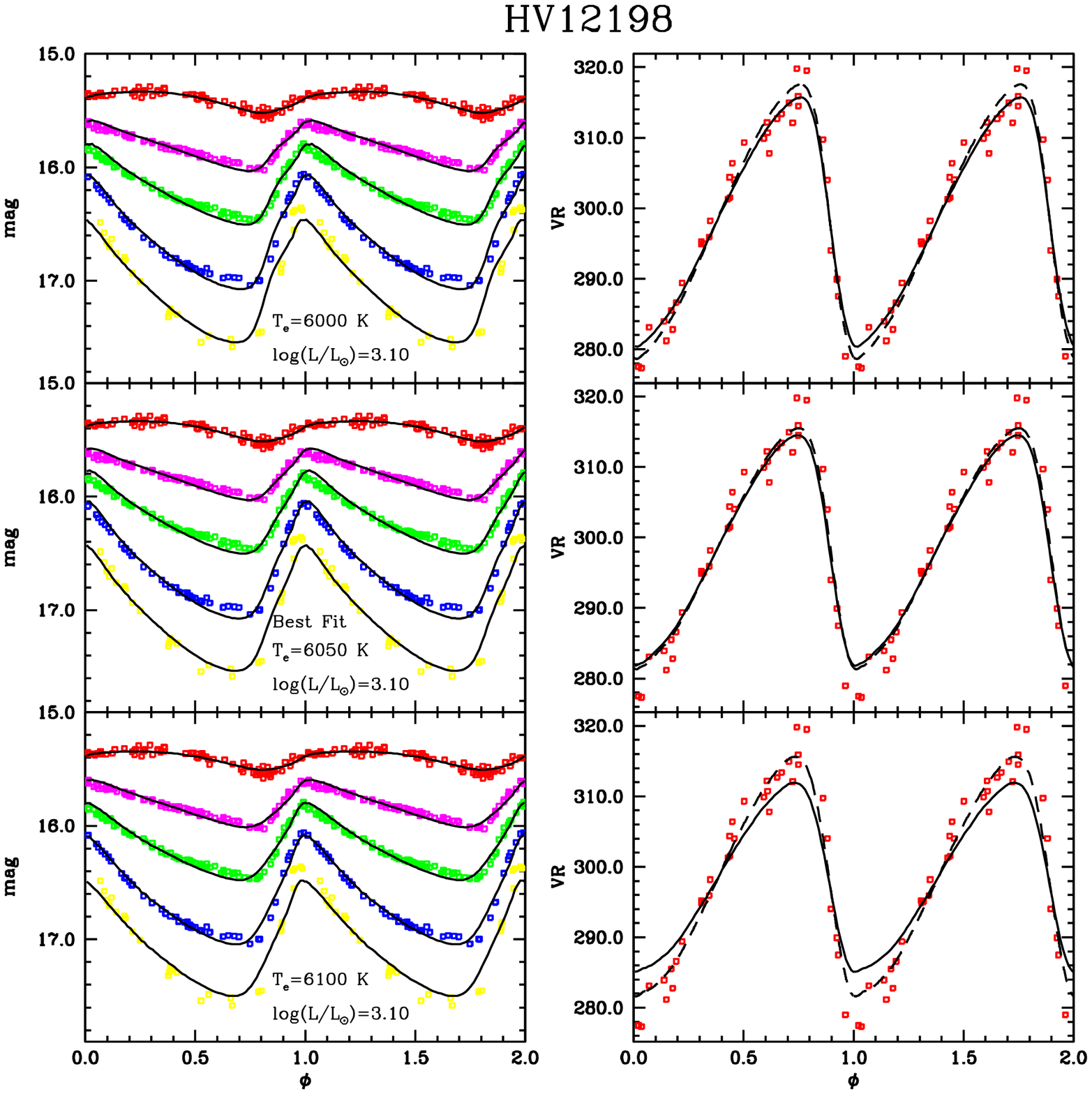}
\caption{The match between model curves and data is shown for the best
  fit (central panel) and other two possible models, characterized by
  the same luminosity of the best fit one but  slightly lower ($T_e$=6000 K
  top panels) and higher ($T_e$=6100 K bottom panels) effective
  temperature. As in the previous 
  figures, the photometry is modeled in the left panels and the
  radial velocity in the right ones. Moreover, the light curves in the U, B,
  V, I and K band are represented from the bottom to the top. In the
  radial velocity panels, the dashed lines represent the 
  model radial velocity curves obtained from the fit with free p factor. }\label{fig-bf-hv12198}               
\end{center}
\end{figure*}

Fig.\ref{fig-bf-hv12198} shows three possible models which
describe the photometric and radial velocity data of HV 12198. The
minimum $\chi^2$ for the photometry is obtained for the model at
$T_e=6100$ K and $\log(L/L_\odot)=3.10$ dex (bottom panel). However,
this does not reproduce 
accurately the amplitude neither of the U band photometry nor of the
radial velocity data. Decreasing the parameter of the convection
efficiency, $\alpha_{ml}$, would produce the simultaneous increment of the
amplitudes of both light and radial velocity curves. In this way we
would recover the amplitudes of both the U band and the radial
velocity curves, but
we would decrease the accuracy of the fit in the remaining bands. As in the case
of HV 12199, the radial velocity amplitude could be reproduced by
decreasing the p factor to the barely acceptable value of $\sim 1$. If we
exclude the U band, the model matching both the photometry and
the radial velocity data  is the one shown in the
top panel of Fig.\ref{fig-bf-hv12198}, corresponding to $T_e$=6000 K
and $\log(L/L_\odot)=3.1$. In this case we obtain an accurate fit of
the radial velocity data with the chosen projection factor value, which
decreases to 1.15 if we consider it as a free parameter in the
$\chi^2$ minimization. On the basis of these results, we decided to
consider as our best fit model for HV 12198 an intermediate case between
those already described. It is shown in the central panel, includes
the U band fitting  and provides a match of the radial velocity curve
which is somewhat intermediate between those of the models in the top
and bottom panels.  

\begin{table*}
\begin{center}
\caption{In the top part of the table adopted structural parameters (with
  uncertainties) of the best fit models obtained for the chemical
  composition Z=0.008 and Y=0.25: (1) Cepheid name, (2-3) mass, (4-5)
  luminosity, (6) canonical luminosity, (7-8) temperature, (9-10)
  barycentric velocity, (11-12) p factor, (13) convection efficiency
  parameter. In the bottom part of the table the distance moduli (with
  uncertainties) for all the photometric band obtained from the fit:
  (1) Cepheid name, (2-3) U band, (4-5) B band, (6-7) V band, (8-9) I
  band, (10-11) K band.}   
\begin{tabular}{c @{} c @{} c @{} c @{} c @{} c @{} c @{} c @{}
    c @{} c @{} c @{} c @{} c}
\hline
\hline
\multicolumn{13}{c}{Structural parameters}\\
\hline
Name &\hspace{0.1cm} $\frac{M}{~M_\odot}$ &\hspace{0.1cm} $\delta
\left(\frac{M}{~M_\odot}\right)$ &\hspace{0.1cm}
$\log\left(\frac{L}{~L_\odot}\right)$ &\hspace{0.1cm} 
$\delta\log\left(\frac{L}{~L_\odot}\right)$ &\hspace{0.1cm}
$\log\left(\frac{L}{~L_\odot}\right)_{can}$ 
&\hspace{0.1cm} T(K) &\hspace{0.1cm} $\delta$T &\hspace{0.1 cm}
$\gamma$ (km/s) &\hspace{0.1cm} $\delta\gamma$ &\hspace{0.1cm}
p& \hspace{0.1cm} $\delta$p &\hspace{0.1cm} $\alpha_{ml}$ \\
\hline
HV 12197 &\hspace{0.1cm} 4.6  &\hspace{0.1cm} $\pm$0.2 &\hspace{0.1cm}
3.045 &\hspace{0.1cm} $\pm$0.012 &\hspace{0.1cm} 3.01 &\hspace{0.1cm}
5950 &\hspace{0.1cm} $\pm$12 &\hspace{0.1cm} 298.3
&\hspace{0.1cm} $\pm$0.9 &\hspace{0.1cm} 1.330 &\hspace{0.1cm}
$^{+0.025}_{-0.003}$ &\hspace{0.1cm} 2.0 \\
HV 12198 &\hspace{0.1cm} 4.2 &\hspace{0.1cm} $\pm$0.1 &\hspace{0.1 cm}
3.10 &\hspace{0.1cm} $\pm$0.01 &\hspace{0.1 cm} 2.88 &\hspace{0.1
  cm} 6050 &\hspace{0.1cm} $\pm$12 &\hspace{0.1cm} 298.57
&\hspace{0.1cm} $\pm$0.09 &\hspace{0.1cm} 1.216 &\hspace{0.1cm}
$^{+0.002}_{-0.102}$ &\hspace{0.1 cm} 2.0 \\
HV 12199 &\hspace{0.1 cm} 3.5 &\hspace{0.1cm} $\pm$0.1  &\hspace{0.1 cm}
2.91 &\hspace{0.1cm} $\pm$0.01 &\hspace{0.1 cm} 2.62 &\hspace{0.1
  cm} 6125 &\hspace{0.1cm} $\pm$12 &\hspace{0.1cm}
300.9 &\hspace{0.1cm} $\pm$1.0 &\hspace{0.1cm}1.17 &\hspace{0.1cm}
$^{+0.03}_{-0.04}$ &\hspace{0.1cm} 2.0 \\ 
We 2 &\hspace{0.1 cm}4.30 &\hspace{0.1cm} $\pm$0.15  &\hspace{0.1
  cm}3.00 &\hspace{0.1cm} $\pm$0.01  &\hspace{0.1 cm} 
2.92 &\hspace{0.1 cm} 5925 &\hspace{0.1cm} $\pm$12 &\hspace{0.1cm}
302.6 &\hspace{0.1cm} $\pm$0.6 &\hspace{0.1cm}1.232 
&\hspace{0.1cm} $^{+0.006}_{-0.012}$ &\hspace{0.1cm} 1.9\\ 
V 6 &\hspace{0.1 cm}4.0 &\hspace{0.1cm} $\pm$0.1  &\hspace{0.1 cm}
3.03 &\hspace{0.1cm} $\pm$0.01  &\hspace{0.1 cm} 
2.81 &\hspace{0.1 cm} 6300 &\hspace{0.1cm} $\pm$12  &\hspace{0.1cm}
300.6 &\hspace{0.1cm} $\pm$1.0 &\hspace{0.1cm} 1.00 &\hspace{0.1cm} :
&\hspace{0.1cm} 1.8 \\   
\hline
\multicolumn{11}{c}{Distance moduli (mag)} \\ 
\hline
Name &\hspace{0.1cm} $\mu_U$  &\hspace{0.1cm} $\delta \mu_U$
&\hspace{0.1cm} $\mu_B$  &\hspace{0.1cm} $\delta \mu_B$ 
&\hspace{0.1cm} $\mu_V$  &\hspace{0.1cm} $\delta \mu_V$ 
&\hspace{0.1cm} $\mu_I$  &\hspace{0.1cm} $\delta \mu_I$ 
&\hspace{0.1cm} $\mu_K$  &\hspace{0.1cm} $\delta\mu_K$\\
\hline
HV 12197 &\hspace{0.1cm} ... &\hspace{0.1cm} ... &\hspace{0.1cm} 19.09
&\hspace{0.1cm} $\pm$0.04 &\hspace{0.1cm} 18.96 &\hspace{0.1cm} $\pm$0.05
&\hspace{0.1cm} 18.89 &\hspace{0.1cm} $\pm$0.06 &\hspace{0.1cm}
18.68 &\hspace{0.1cm} $\pm$0.04 \\ 
HV 12198 &\hspace{0.1cm} 19.15 &\hspace{0.1cm} $\pm$0.08 &\hspace{0.1cm}
 19.13 &\hspace{0.1cm} $\pm$0.05 &\hspace{0.1cm} 18.98 &\hspace{0.1cm}
 $\pm$0.05 &\hspace{0.1cm} 18.81 &\hspace{0.1cm} $\pm$0.04
 &\hspace{0.1cm} 18.60 &\hspace{0.1cm} $\pm$0.03 \\ 
HV 12199 &\hspace{0.1cm} 18.97 &\hspace{0.1cm} $\pm$0.05
&\hspace{0.1cm}18.96 &\hspace{0.1cm} $\pm$0.07 &\hspace{0.1cm} 18.83
&\hspace{0.1cm} $\pm$0.06 &\hspace{0.1cm} 18.68 &\hspace{0.1cm}
$\pm$0.04 &\hspace{0.1cm} 18.49 &\hspace{0.1cm} $\pm$0.04 \\
We 2 &\hspace{0.1cm} 18.70 &\hspace{0.1cm} $\pm$0.08 &\hspace{0.1cm}
18.82 &\hspace{0.1cm} $\pm$0.05 &\hspace{0.1cm} 18.78 &\hspace{0.1cm}
$\pm$0.04 &\hspace{0.1cm} 18.71 &\hspace{0.1cm} $\pm$0.03
&\hspace{0.1cm} 18.53 &\hspace{0.1cm} $\pm$0.06 \\
V6 &\hspace{0.1cm} 19.10 &\hspace{0.1cm} $\pm$0.04 &\hspace{0.1cm}
19.19 &\hspace{0.1cm} $\pm$0.02 &\hspace{0.1cm} 19.03 &\hspace{0.1cm}
$\pm$0.02 &\hspace{0.1cm} 18.87 &\hspace{0.1cm} $\pm$0.02 &\hspace{0.1cm}
 18.61 &\hspace{0.1cm} $\pm$0.07 \\
\hline
\hline
\end{tabular}
\label{tab-params}
\end{center}
\end{table*}

\subsection{Comparison with the spectroscopic data}\label{sec-spec-data}
\begin{figure}
\begin{center}
\includegraphics[width=84mm]{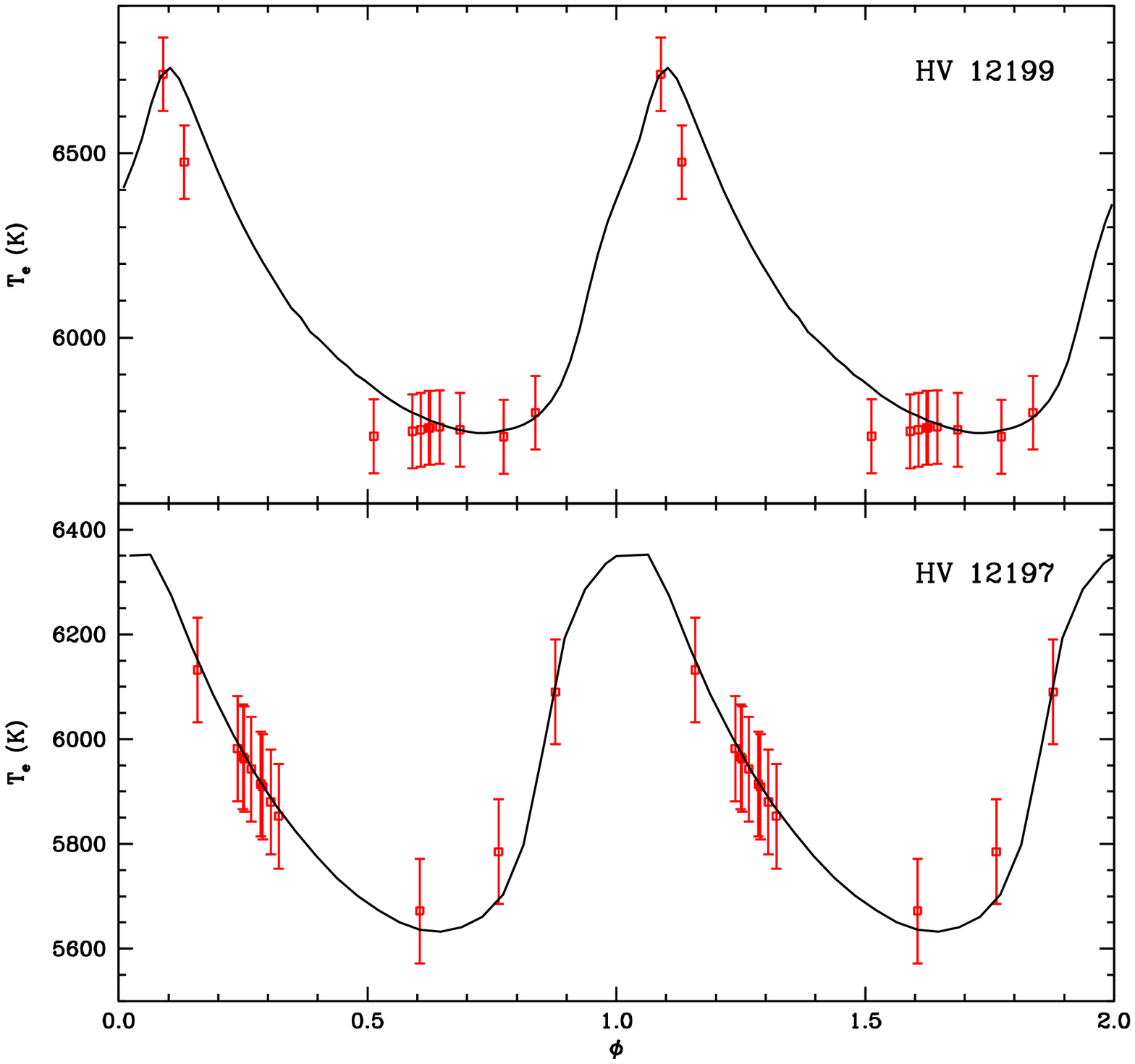}
\caption{Effective temperature curves predicted by models (solid
  lines) with overplotted spectroscopic determinations by
  \citet{muc11} (empty squares) for HV 12197 and HV 12199 in the
  bottom and top panels respectively. }\label{fig-Tspec-Tmod}        
\end{center}
\end{figure}

As a test for the accuracy of our procedure, we compared the effective
temperature  predicted by the best fit models for HV 12197 and HV
12199, with the results obtained by \citet{muc11} using an independent
spectroscopic determination. The two panels of
Fig.\ref{fig-Tspec-Tmod} show the model temperature as a function of
the pulsational phase of the two quoted Cepheids and their
spectroscopic temperatures, given by \citet{muc11} 
with a typical error bar of 100 K.   

It is evident that models reproduce the spectroscopic data with great
accuracy, thus further supporting the predictive capabilities of the 
adopted theoretical scenario, as well as of the model fitting technique. 

\subsection{Distance and reddening}\label{sec-dist}
Beyond the intrinsic stellar
parameters of the best fit models (chemical composition, stellar mass,
effective temperature,
luminosity and convective efficiency parameter) and the corresponding
projection factor, Tab.\ref{tab-params} 
reports the resulting distance moduli in all the observed bands, $\mu_i$
(i=U, B, V, I, K), with the exception of HV 12197, for which the U band data
consists only of three measurements and is not used to infer the
distance. The range of values we find for the apparent distance moduli
of the selected stars can be at least in part understood in terms of
differential reddening, but other effects might in principle be at work
(see below). 
Using the obatined apparent distance moduli and the  
photometric band effective wavelengths, $\lambda_i$, we fitted the
\citet{car89} extinction law to derive simultaneously the absorption,
$A_V$, and the true distance modulus, $\mu_0$, for the
selected Cepheids. To this aim, we minimized the following 
$\chi^2$ function by varying the two unknown parameters:  
\begin{equation}
\chi^2=\sum^{N_{bands}}_{i=1}\left [ \mu_i -
  \left(a(x_i)+\frac{b(x_i)}{R_V}\right )A_V - \mu_0 \right ]
\end{equation}
where the total to selective extinction ratio is fixed to $R_V=3.3$
\citep{fea87}, $x_i\equiv \frac{1}{\lambda_i}$ is the inverse of the
i band effective wavelength and the expressions for $a(x)$ and $b(x)$
are defined in \citet{car89}.  

\begin{figure*}
\begin{center}
\includegraphics{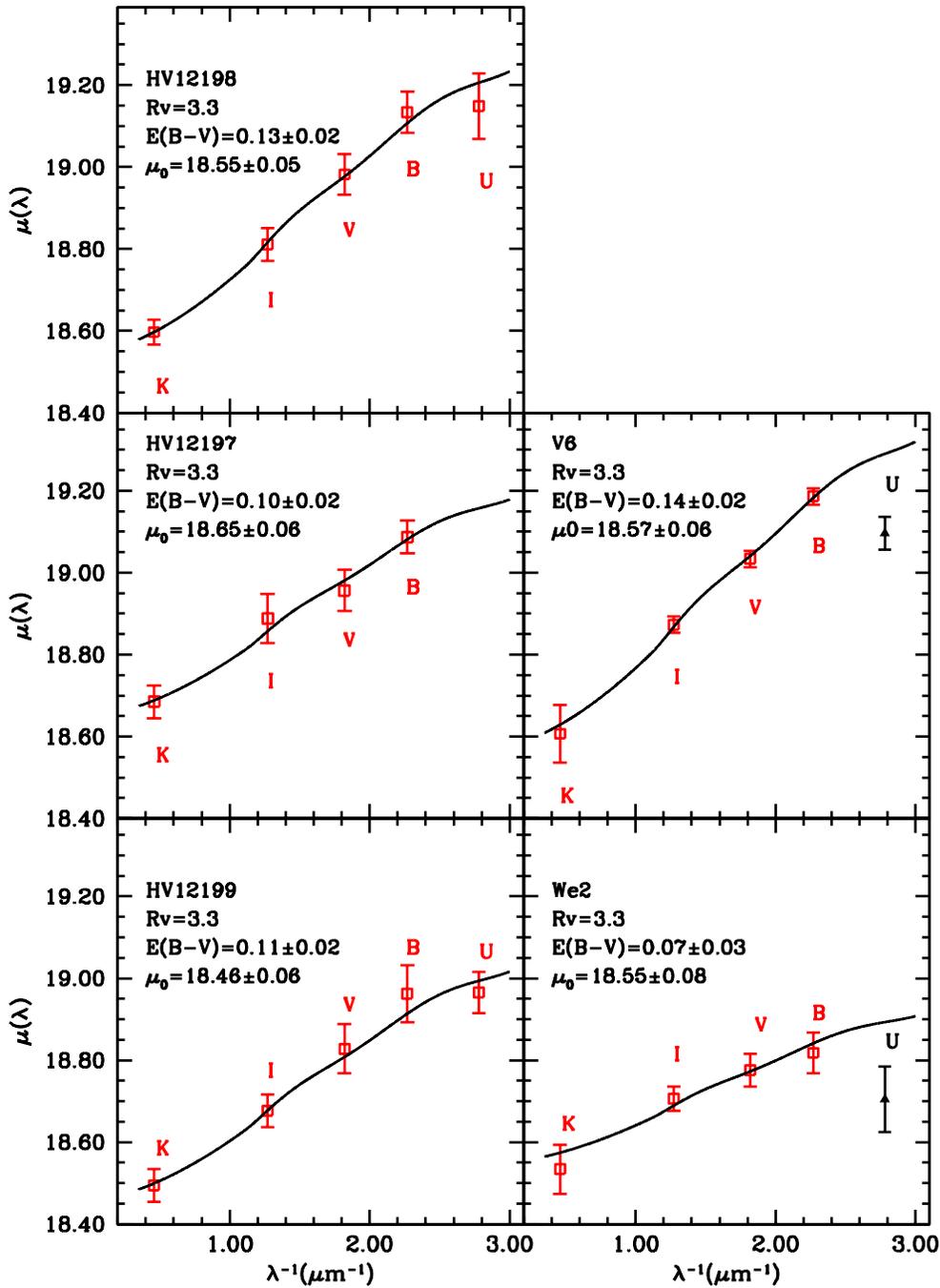}
\caption{The result of the best fit of the \citet{car89} reddening
  law is plotted for the five selected Cepheids (solid line). The data
points (empty squares) represents the distance moduli obtained from
the model fitting in the photometric bands considered in this
work. The two filled triangles represent the U band distance moduli of
V 6 and We 2 excluded from the fit.}\label{fig-fitCardelli89}         
\end{center}
\end{figure*}

The results of the fit are shown in fig.~\ref{fig-fitCardelli89} and the
best fit parameters are listed in Tab.~\ref{tab-parCar89}.
To be conservative, the errors on the individual band distance moduli
include the rms of the fitting procedure and the error related to the
selection of  the best fit model\footnote{Half the difference
between the distance moduli obtained from
the ``secondary'' models defined above}, summed in quadrature.
Finally, the uncertainties on the parameters of the Cardelli fit have
been derived from the $\chi^2$ confidence level at
$1\sigma$. In the fitting procedure, we decided to exclude the U band
distance moduli of We 2 and V 6 because they significantly deviate from
the expected trend  (see the filled triangles in 
fig.\ref{fig-fitCardelli89}).  

Inspection of Tab.~\ref{tab-parCar89}, suggests
color excess estimates larger than the
typically adopted value for NGC 1866,  namely E(B-V)=0.06 mag \citep[][and
  references therein]{sto05,mol12}, with the exception of the result for
  We 2.  In fact, the resulting weighted mean value is  E(B-V)=0.11$\pm$0.01
mag, consistent with the result by 
\citet{gro03},  E(B-V)=0.12$\pm$0.02 mag, as based on the simultaneous
fit of the NGC 1866 Cepheid Period--Luminosity relation, in the B,
V and I photometric bands. 

As for the distance modulus of NGC 1866, 
a weighted mean of the obtained results for the five stars, provides
$\mu_0=18.56\pm0.03$ mag.
 Here, the uncertainty is only statistical but we are aware that
  several systematic effects are at work when reproducing observing
  quantities with pulsation models. First, we have to consider the
  effect of the adopted physical assumptions, namely the equation of
  state and the opacity tables. Previous theoretical investigations
   \citep{petroni03,valle09} show that the effect of varying these ingredients is marginal and dominated by the
  effect of the model spatial risolution. The latter affects the
  predicted pulsation amplitudes and in turn the intrinsic parameters
  of the obtained best fit models.  In particular, increasing the
  adopted spatial resolution by 10 mesh zones can imply a variation of about
  100 K in the predicted effective temperature and of few hundredths of
  dex in the predicted luminosity level.  
Another important source of uncertainty
is the treatment of the pulsation and convection coupling. Even if we
adopt a nonlinear nonlocal time-dependent treatment of convection
\citep[see][for details]{s82,bms99}, the equation system is closed by
adopting a free parameter related to the mixing lenght. Variations of
the mixing lenght affect the pulsation efficiency and amplitude. In
particular, by
varying the mixing lenght parameters by more than $\pm0.05$ from the value reported in Tab.\ref{tab-params},
 we are not able to  reproduce the observed curves, with resulting
changes in the predicted luminosity levels smaller than $\pm0.03$ dex.
Obviously, this is only the effect of varyingh the mixing lenght
parameter,  but within the same turbulent convective model.  Assuming a
different treatment of convection might in principle produce larger errors even if
we consider quite encouraging that the application of the model fitting
technique, from different groups
\citep[e.g.][]{bcm02,kw02,mc05,kw06,mcn07} and using different approaches
to the treatment of the
pulsation-convection coupling, gives consistent results for the LMC
distance modulus \citep{mc05,m09}. 
Finally, for what concern the light curves we have to consider the
uncertainty  on the adopted model atmosphere in transforming
bolometric into B,V,I,K variations.
According to our previous experience we know that theoretical
predictions are dependent on the set of static atmosphere models
adopted for transforming temperatures into colors.  For example, changing the
adopted model atmospheres from \citet{cgk97a,cgk97b} to \citet{k93} 
 produces colour effects of the order of 0.01 mag.
In conclusion, to be conservative, we assume a systematic effect of
$\pm$ 0.1 mag on the inferred distance modulus, as due to all the
above mentioned
theoretical uncertainties. On this basis our final estimate of NGC1866
distance modulus is $\mu_0=18.56\pm0.03 (stat) \pm0.1 (syst)$ mag.

 This result is in agreement within the
uncertainty interval with the value $18.51\pm0.03$ mag,  obtained by
\citet{mol12} from the Baade--Wesselink method, using the same p
factor adopted as reference value in this work. Their
estimates of the distance to HV 12197, We 2 and HV 12198, are in
excellent agreement with the values reported in
Tab.~\ref{tab-parCar89}. In particular, 
they found 18.63$\pm$0.12 mag, 18.54$\pm$0.09 mag and
18.59$\pm$0.08 mag for HV 12197, We 2 and HV 12198 respectively. For
the remaining two stars, HV 12199 and V 6, they 
found 18.62$\pm$0.10 mag and 18.83$\pm$0.11 mag, respectively,  both
systematically longer than our estimates, although consistent with them
within the errors. 
 In a recent work, using the infrared surface brightness method,
\citet{sto11b} obtained the distance for a sample of Cepheids in the
LMC, including HV 12197, HV 12199 and HV12198, and their final value (18.45 $\pm$ 0.04) is in agreement with
our results within the errors. 
However,  they used a period dependent p factor given by the equation
$p=1.550(\pm0.04)-0.186(\pm0.06)\log P$. This is required to obtain
distances to LMC Cepheids  independent of their pulsation periods and
distances to Galactic Cepheids in agreement with the HST parallaxes, with
their surface brightness method \citep[see][]{sto11a}. As stated by
the authors themeselves, this relation is not easily reconciled with
recent theoretical work (e.g. Nardetto et al. 2009) and provides
p-factor values for  short-period Cepheids (not less than 1.4),
significantly larger than the results derived in the present study. 
This discrepancy could be due, at least in part, to limitations of our
treatment of the coupling between pulsation and convection (see
discussion above) but we have to note that the debate on the p factor
and on its possible dependence on the pulsation period
is still open in the recent literature \citep[see e.g.][and references
therein]{ngeow12}.
 
\begin{table}
\begin{center}
\caption{Parameters obtained by fitting Cardelli extinction law:
  the Cepheids are listed in the first column, the second and third
  columns contain, respectively, the extinction, $A_V$, and the
  reddening, $E(B-V)$, the intrinsic distance modulus, $\mu_0$ is in
  the last column.}  
\begin{tabular}{c @{} c @{} c @{} c}
\hline
\hline
Name &\hspace{0.1 cm}$A_V$ (mag)&\hspace{0.1 cm} E(B-V) (mag)&\hspace{0.1cm}
$\mu_0$ (mag)\\
\hline
HV 12197 &\hspace{0.1 cm}$0.33\pm0.07$ &\hspace{0.1 cm}$0.10\pm0.02$
&\hspace{0.1 cm} $18.65\pm0.06$ \\ 
HV 12198 &\hspace{0.1 cm}$0.43\pm0.07$ &\hspace{0.1 cm}$0.13\pm0.02$
&\hspace{0.1 cm} $18.55\pm0.05$ \\ 
HV 12199 &\hspace{0.1 cm}$0.37\pm0.07$ &\hspace{0.1 cm}$0.11\pm0.02$
&\hspace{0.1 cm} $18.46\pm0.06$ \\ 
We 2 &\hspace{0.1 cm}$0.23\pm0.10$ &\hspace{0.1 cm}$0.07\pm0.03$
&\hspace{0.1 cm} $18.55\pm0.08$ \\ 
V 6 &\hspace{0.1 cm}$0.46\pm0.07$ &\hspace{0.1 cm}$0.14\pm0.02$
&\hspace{0.1 cm} $18.57\pm0.06$ \\ 
\hline
\hline
\end{tabular}
\label{tab-parCar89}
\end{center}
\end{table}

\subsection{Mass--Luminosity relation}\label{sec-ML-relation}

The obtained stellar masses for the Cepheids in NGC 1866, as reported
in Tab.\ref{tab-params},  cover a range
of values that might be the signature of differential
mass loss. This occurrence is in agreement with previous findings by \citet[][]{br04}.
Finally, we compared the derived masses and luminosities, as reported in
Tab.\ref{tab-params}, with an evolutionary Mass--Luminosity relation  (MLR)
\citep{bon00}, either neglecting or including mild overshooting
according to the prescriptions by \citet{chi93}.

\begin{figure}
\begin{center}
\includegraphics[width=84mm]{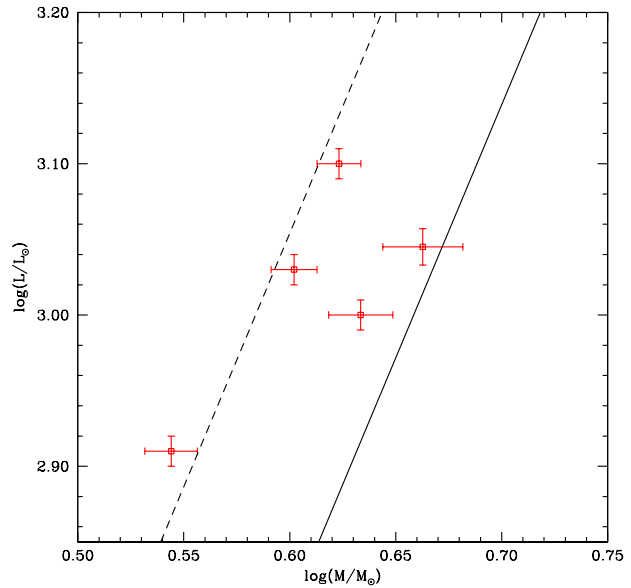}
\caption{The masses and luminosities of the analyzed Cepheids (empty
  squares) are compared with the canonical Mass--Luminosity relation
  (solid line) and with the mild overshooting Mass--Luminosity
  relation (dashed line).}\label{fig-ml-relation}         
\end{center}
\end{figure}

Fig.\ref{fig-ml-relation} clearly shows that the analyzed Cepheids do
not follow one of the two relations, but they are randomly placed at
intermediate luminosities between those predicted by the canonical and
the mild overshooting MLRs, with the exception of the HV 12199
which results to be slightly more luminous than the mild overshooting
prediction, although consistent with it within the uncertainties.

The fact that different luminosities are predicted for a given mass
might suggest that the investigated Cepheid do not follow a
MLR but are instead stochastically affected by
some noncanonical phenomenon, likely a combination of mild
overshooting and mass loss. However the application of the method to a
larger sample of pulsators is needed in order to draw any reliable
conclusion on the cause of the overluminosity distribution with
respect to the canonical one.

\section{Conclusions}\label{sec-conclusions}
We have used nonlinear convective pulsation models computed by our
team to reproduce the multifilter (U,B,V, I and K)  light and radial velocity curves of five Classical
Cepheids in NGC 1866, a young massive cluster of the Large Magellanic
Cloud.  The resulting best fit models give us information on the
intrinsic stellar parameters and the individual distances of the
investigated Cepheids. In the case of HV 12197 and HV 12199 the
obtained effective 
temperature and its variation with the pulsation phase has been found
to be in very good agreement with the spectroscopic determinations
within their uncertainties.
The masses and luminosities, obtained for all the five investigated
pulsators, from this model fitting technique
satisfy  a slightly brighter
Mass--Luminosity relation than the canonical evolutionary one,
indicating that noncanonical phenomena such as  mild
overshooting and/or mass loss are at work. As for the inferred distances, the
individual values have been found to be consistent with each other within the
uncertainties. Moreover, their weighted mean value corresponds to a distance
modulus of 18.56$\pm$0.03 (stat) $\pm$ 0.1 (syst) mag, in agreement with several independent
results in the literature. In particular the obtained
result for the distance to NGC1866 is in excellent agreement with the
results obtained from the application of the model fitting technique to LMC
field Cepheids, RR Lyrae and $\delta$ Scuti variables
\citep[][]{bcm02,mc05,mcn07}. 

\section{Acknowledgments}
This paper utilizes public domain data obtained by the MACHO Project,
jointly funded by the US Department of Energy through the University
of California, Lawrence Livermore National Laboratory under contract
No. W-7405-Eng-48, by the National Science Foundation through the
Center for Particle Astrophysics of the University of California under
cooperative agreement AST-8809616, and by the Mount Stromlo and Siding
Spring Observatory, part of the Australian National University.
We thank an anonymous referee for helpful comments and suggestions.



\begin{thebibliography}{99}
\bibitem[Arp 
\& Thackeray(1967)]{at67} Arp, H., \& Thackeray, A.~D.\ 1967, ApJ,
149, 73 
\bibitem[Barmina et 
al.(2002)]{ba02} Barmina, R., Girardi, L., \& Chiosi, C.\ 2002, A\&A, 385, 847 
\bibitem[Bessell \& Brett(1988)]{bes88} Bessell, M.S. \& Brett, J.M.,
  1988, PASP, 100, 1134 
\bibitem[Bono et al.(2000a)]{bcm00} Bono, G., Castellani, V., 
\& Marconi, M., 2000, ApJ, 529, 293
\bibitem[Bono et al.(2000b)]{bon00} Bono, G., Caputo, F., 
Cassisi, S., et al.\ 2000, ApJ, 543, 955 
\bibitem[Bono et al.(2002)]{bcm02} Bono, G., Castellani, V., 
\& Marconi, M.\ 2002, ApJL, 565, L83
\bibitem[Bono \& Stellingwerf(1994)]{bs94} Bono, G., \& Stellingwerf,
  R.~F.\ 1994, ApJS, 93, 233  
\bibitem[Bono et al.(1999)]{bms99} Bono, G., Marconi, M., 
\& Stellingwerf, R.~F.\ 1999, ApJS, 122, 167 
\bibitem[Brocato et al.(1989)]{br89} Brocato, E., Buonanno, 
R., Castellani, V., \& Walker, A.~R.\ 1989, ApJS, 71, 25 
\bibitem[Brocato et 
al.(1994)]{br94} Brocato, E., Castellani, V., \& Piersimoni, A.~M.\ 1994, A\&A, 290, 59 
\bibitem[Brocato et al.(2003)]{br03} Brocato, E., 
Castellani, V., Di Carlo, E., Raimondo, G., 
\& Walker, A.~R.\ 2003, AJ, 125, 3111 
\bibitem[Brocato et al.(2004)]{br04} Brocato, E., Caputo, 
F., Castellani, V., Marconi, M., \& Musella, I.\ 2004, AJ, 128, 1597 
\bibitem[Buchler  
\& Szab{\'o}(2007)]{busz07} Buchler, J.~R., \& Szab{\'o},
R.\ 2007, ApJ, 660, 723 
\bibitem[Cardelli et al.(1989)]{car89} Cardelli, J.A., 
Clayton, G.C., \& Mathis, J.S., 1989, ApJ, 345, 245
\bibitem[Castelli et 
al.(1997a)]{cgk97a} Castelli, F., Gratton, R.~G., \& Kurucz, R.~L.\ 1997a, A\&A, 324, 432 
\bibitem[Castelli et 
al.(1997b)]{cgk97b} Castelli, F., Gratton, R.~G., \& Kurucz, R.~L.\ 1997b, A\&A, 318, 841 
\bibitem[Chiosi et 
al.(1989)]{c89} Chiosi, C., Bertelli, G., Meylan, G., \& Ortolani, S.\ 1989, A\&A, 219, 167
\bibitem[Chiosi et al.(1993)]{chi93} Chiosi, C., Wood, P.R., 
\& Capitanio, N., 1993, ApJS, 86, 541
\bibitem[Di Criscienzo et al.(2004)]{dic04} Di Criscienzo, 
M., Marconi, M., \& Caputo, F.\ 2004, ApJ, 612, 1092 
\bibitem[Di Fabrizio et al.(2002)]{dif02} Di Fabrizio, L., 
Clementini, G., Marconi, M., et al.\ 2002, MNRAS, 336, 841 
\bibitem[Feast \& Walker(1987)]{fea87} Feast, M.W. \& Walker, A.R.,
  1987, ARA\&A, 25, 345 
\bibitem[Feast \& Catchpole(1997)]{fea97} Feast, M.W. \&, Catchpole,
  R.M., 1997, MNRAS, 286, L1 
\bibitem[Feuchtinger(1999)]{feu99} Feuchtinger, M.~U.\ 1999, A\&AS, 136, 217
\bibitem[Fiorentino et al.(2007)]{f07} Fiorentino, G., Marconi, M., Musella, I., \& Caputo, F.\ 2007, A\&A, 476, 863 
\bibitem[Fischer et al.(1992)]{fi92} Fischer, P., Welch, 
D.~L., Cote, P., Mateo, M., \& Madore, B.~F.\ 1992, AJ, 103, 857 
\bibitem[Freedman(1988)]{fre88} Freedman, W.L. 1988, ApJ, 326, 691 
\bibitem[Freedman et al.(2001)]{f01} Freedman, W.~L., 
Madore, B.~F., Gibson, B.~K., et al.\ 2001, ApJ, 553, 47 
\bibitem[Gieren et al.(1994)]{g94} Gieren, W.~P., Richtler, 
T., \& Hilker, M.\ 1994, ApJL, 433, L73 
\bibitem[Gieren et al.(2000)]{gie00} Gieren, W.P., G{\'o}mez, M.,
  Storm, J., et al., 2000, ApJS, 129, 111   
\bibitem[Groenewegen \& Salaris(2003)]{gro03} Groenewegen, M.A.T., \&
  Salaris, M., 2003, A\&A, 410, 887 
\bibitem[Keller \& Wood(2002)]{kw02} Keller, S.~C., \& Wood, P.~R.\
2002, Apj, 578, 144 
\bibitem[Keller \& Wood(2006)]{kw06} Keller, S.~C., \& Wood, P.~R.\
  2006, Apj, 642, 834 
\bibitem[Kurucz(1993)]{k93} Kurucz, R.~L.\ 1993, Physica 
Scripta Volume T, 47, 110 
\bibitem[Labhardt, Sandage \& Tammann(1997)]{lab97} Labhardt, L.,
  Sandage, A., Tammann, G.A., 1997, A\&A, 322, 751
\bibitem[Leavitt \& Pickering(1912)]{l1912} Leavitt, H.~S., \& Pickering, E.~C.\ 1912, Harvard College Observatory Circular, 173, 1 
\bibitem[Madore \& Freedman(1991)]{mf91} Madore, B.~F., \& Freedman, W.~L.\ 1991, PASP, 103, 933 
\bibitem[Marconi \& Clementini(2005)]{mc05} Marconi, M., \& Clementini,
G.\ 2005, AJ, 129, 2257
\bibitem[Marconi \& Degl'Innocenti(2007)]{md07} Marconi, M., \&
Degl'Innocenti, S.\ 2007, A\&A, 474, 557
\bibitem[Marconi(2009)]{m09} Marconi, M.\ 2009, Mem. SAIt, 80, 141 
\bibitem[Marconi et al.(2010)]{m10} Marconi, M., Musella, 
I., Fiorentino, G., et al.\ 2010, ApJ, 713, 615 
\bibitem[McNamara et al.(2007)]{mcn07} McNamara, D.~H.,
Clementini, G., \& Marconi, M.\ 2007, AJ, 133, 2752
\bibitem[Molinaro et al.(2012)]{mol12} Molinaro, R., et al., 2012,
  ApJ, 748, 69 
\bibitem[Mucciarelli et al.(2011)]{muc11} Mucciarelli, A., 
Cristallo, S., Brocato, E., et al., 2011, MNRAS, 413, 837
\bibitem[Musella et al.(2006)]{mus06} Musella, I., et al., 2006,
  Mem. SAIT, 77, 291
 \bibitem[Nardetto et al.(2009)]{nar09} Nardetto, N., Gieren, W., Kervella,
   P., et al., 2009, A\&A, 502, 951
\bibitem[Natale et al.(2008)]{nat08} Natale, G., Marconi, M., 
\& Bono, G., 2008, ApJL, 674, L93
\bibitem[Ngeow et 
al.(2012)]{ngeow12} Ngeow, C.-C., Neilson, H.~R., Nardetto, N., \& Marengo, M.\ 2012, A\&A, 543, A55 
\bibitem[Olivier 
\& Wood(2005)]{ow05} Olivier, E.~A., \& Wood, P.~R.\ 2005, MNRAS, 362, 1396  
\bibitem[Petroni et al.(2003)]{petroni03} Petroni, S., Bono, G., 
Marconi, M., \& Stellingwerf, R.~F.\ 2003, ApJ, 599, 522 
\bibitem[Riess et al.(2011)]{rie11} Riess, A.~G., Macri, L., 
Casertano, S., et al.\ 2011, ApJ, 730, 119 
\bibitem[Riess et al.(2012)]{rie12} Riess, A.~G., Macri, L., 
Casertano, S., et al.\ 2012, ApJ, 752, 76 
\bibitem[Robertson(1974)]{r74} Robertson, J.~W.\ 1974, 
ApJ, 191, 67 
\bibitem[Saha et al.(2001)]{s01} Saha, A., Sandage, A., 
Tammann, G.~A., et al.\ 2001, ApJ, 562, 314 
\bibitem[Stellingwerf(1982)]{s82} Stellingwerf, R.~F.\ 
1982, ApJ, 262, 330 
\bibitem[Storm et al.(2004)]{sto04} Storm, J., et al, 2004, A\&A, 415, 521
\bibitem[Storm et al.(2005)]{sto05} Storm, J., Gieren, W.P., Fouqu\'e,
  P., Barnes III, T.G. \& G\'omez, M., 2005, A\&A, 440, 487 
\bibitem[Storm et al.(2011)]{sto11a} Storm, J., Gieren, W., Fouqu{\'e},
  P., et al., 2011, A\&A, 534, A95 
\bibitem[Storm et al.(2011)]{sto11b} Storm, J., Gieren, W.,
  Fouqu{\'e}, P., et al.\ 2011, A\&A, 534, A95 
\bibitem[Tammann 
\& Reindl(2012)]{tr12} Tammann, G.~A., \& Reindl, B.\ 2012, ApSS, 43 
\bibitem[Testa et al.(1999)]{t99} Testa, V., Ferraro, 
F.~R., Chieffi, A., et al.\ 1999, AJ, 118, 2839 
\bibitem[Testa et al.(2007)]{tes07} Testa, V., et al., 2007,
  A\&A, 462, 599
\bibitem[Udalski et al.(1999)]{u99} Udalski, A., Soszynski, 
I., Szymanski, M., et al.\ 1999, ActaA, 49, 223 
\bibitem[Valle et 
al.(2009)]{valle09} Valle, G., Marconi, M., Degl'Innocenti, S., \&
Prada Moroni, P.~G.\ 2009, A\&A, 507, 1541 
\bibitem[Walker(1995)]{w95} Walker, A.~R.\ 1995, AJ, 110, 
638 
\bibitem[Walker et al.(2001)]{w01} Walker, A.~R., Raimondo, 
G., Di Carlo, E., et al.\ 2001, ApJL, 560, L139 
\bibitem[Wood et al.(1997)]{was97} Wood, P.~R., Arnold, A., 
\& Sebo, K.~M.\ 1997, ApJL, 485, L25 
\bibitem[Walker(2011)]{w11} Walker, A.~R.\ 2011, ApSS, 746 
\bibitem[Welch et al.(1991)]{wel91} Welch, D.L., Mateo, M.,
  C\^{o}t\'e, P., Fischer, P. \& Madore, B.F., 1991, AJ, 101, 490  
\bibitem[Welch \& Stetson(1993)]{ws93} Welch, D.~L., \& Stetson,
P.~B.\ 1993, AJ, 105, 1813 
\end{thebibliography}
\end{document}